\documentclass[12pt]{article}
\usepackage{color, graphicx}
\usepackage{verbatim,amsmath, amssymb, booktabs, ulem}
\usepackage{listings}
\usepackage[numbers,sort&compress]{natbib}
\usepackage[colorlinks=true, linkcolor=blue, filecolor=blue, urlcolor=blue, citecolor=blue, plainpages=false]{hyperref}
\usepackage{natbibspacing}
\usepackage{epsfig}
\usepackage{comment}
\title{A novel approach to the study of conformality in the $SU(3)$
       theory with multiple flavors.}
\author{Richard Brower$^{1,2}$, Anna Hasenfratz$^3$, Claudio Rebbi$^{1,2}$,  \\
	Evan Weinberg$^1$, Oliver Witzel$^2$\footnote{Present address: School of Physics \& Astronomy, The University of Edinburgh, EH9 3FD, UK}   \\
       \small{$^1$Department of Physics, Boston University, Boston, Massachusetts 02215, USA}\\
       \small{$^2$ Center for Computational Science, Boston University, Boston, Massachusetts 02215, USA}\\
        \small{$^3$ Department of Physics, University of Colorado, Boulder, CO 80309, USA} \\
	}

\date{\today}

\begin{document}

\maketitle

\begin{abstract}
We investigate the transition between spontaneous chiral symmetry breaking 
and conformal behavior in the $SU(3)$ theory with multiple fermion
flavors.  We propose a new strategy for studying this
    transition. Instead of changing the number of flavors, we lift the
    mass of a subset of the fermions, keeping the rest of the
   fermions near to the massless chiral limit in order to probe the
    transition. 

\smallskip
Dedicated to the 60$^{\rm th}$ birthday of Academician Valery Rubakov.
\end{abstract}

\section{Introduction}

\label{Introduction}

The discovery of the Higgs meson has put in place the 
final piece of the electroweak sector of the Standard Model~\cite{Aad:2012tfa,Chatrchyan:2012ufa}.  But, contrary to a theory
like QCD where asymptotic freedom guarantees that the ultraviolet
cutoff can be removed, the theory of a self-interacting scalar
 needs an ultraviolet completion.
The theoretically potentially viable options, compatible with present experimental limits fall into two main categories: supersymmetric and composite Higgs models.  Composite Higgs models are based on some new strongly coupled but chirally broken gauge-fermion sector where the Higgs particle is  a fermionic $0^{++}$ bound state. The idea of a composite model was first introduced as ``Technicolor''~\cite{Weinberg:1979bn,Susskind:1978ms},
which was later generalized to ``Extended Technicolor''~\cite{Dimopoulos:1979es,Eichten:1979ah} to accommodate
a mechanism for generating fermion masses.  Rather soon it became clear that  simple ``scaled-up QCD'' models cannot satisfy electroweak phenomenological constraints. ``Walking Technicolor'', a model where 
the underlying coupling constant evolves very slowly, was proposed to remedy the situation~\cite{PhysRevLett.56.1335,Appelquist:1991nm}.
The notion
of walking technicolor was particularly important because
it brought to light the relevance  that the proximity
of an infrared fixed point could play  for strongly coupled
theories of electroweak symmetry breaking.

A general feature of theories of electroweak symmetry breaking based
on strong dynamics is that they predict a variety of new composite
states.  However,  to date, no additional states have
been found at the LHC other than the Higgs particle.  Thus, any
strong dynamics model of electroweak symmetry breaking faces
 the challenge of predicting the existence of a bound scalar
state with the mass and properties of the Higgs, while  the other
states of the system
must be significantly heavier.  In passing, it is worth noting that even the occurrence of a Higgs-like scalar
was not a given in the original formulation of such theories.
Indeed one of the motivation for their introduction was to provide electroweak symmetry breaking in the absence of any Higgs
particle in the low energy spectrum.

\begin{figure}[!ht]
\begin{center}
    \includegraphics[width=8cm]{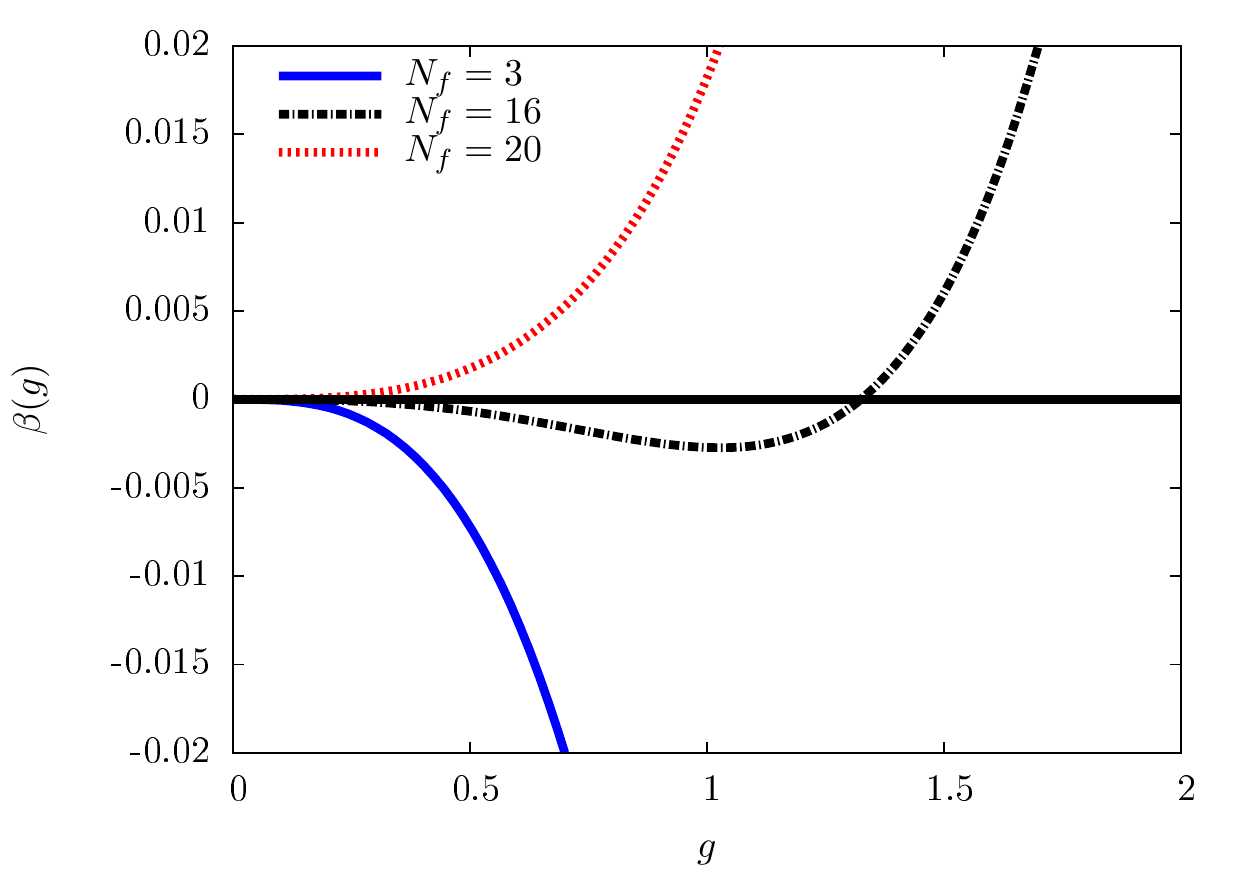} 
\end{center}
    \caption{Two-loop perturbative $\beta$ function for $SU(3)$ with different number of flavors, indicating asymptotically free, fixed point, and IR-free behavior.}
    \label{f1}
\end{figure}

We note that in QCD, the $0^{++}$ $\sigma(550)$ resonance is very close to other hadronic excitations. A phenomenologically viable model must exhibit some new phenomena that separates the $\sigma$ from the rest of the non-Goldstone spectrum.
One possibility which is currently getting a lot of attention
is a theory which exhibits near conformal symmetry.
The presence of a softly broken conformal symmetry might give
origin to a low mass scalar, possibly as a pseudo-dilaton state, while all other composites states
would appear at much higher mass.  This could happen if the theory
is close to  an infrared fixed point as suggested in walking Technicolor scenarios.

In the generalization of QCD to $SU(N_c )$ theories
with $N_f$ fundamental flavors the two-loop $\beta$ function 
is given by
\begin{equation}
\beta(g)=-\beta_0g^3-\beta_1g^5+O(g^7)
\label{eq1}
\end{equation}
with
\begin{equation}
\beta_0=\frac{\Big[\frac{11}{3}\,N_c-\frac{2}{3}\,N_f\Big]}{(4\pi)^2}, \qquad
\beta_1=\frac{\Big[\frac{34}{3}\,N_c^3-\Big(
\frac{34}{3}N_c^2-\frac{1}{N_c}\Big)\,N_f\Big]}{(4\pi)^4}.
\label{eq2}
\end{equation}
When the number of flavors is small both $\beta_0$ and $\beta_1$ are negative, indicative of QCD-like asymptotic freedom. As equations~\ref{eq1} and \ref{eq2} show,  with increasing number of flavors first $\beta_1$ changes sign at $N_f^{(c)}$ and the $\beta$ function  develops a non-trivial zero at some $g=g_*$ value,  then   at $N_f'>N_f^{(c)}$ the value of  $\beta_0$ changes sign as well and  asymptotic 
freedom is lost. Figure~\ref{f1} illustrates this behavior
for $N_c=3$.
For $N_f^{(c)}<N_f<N_f'$  the two-loop
$\beta$ function suggests the presence of an
infrared fixed-point at some value $g_*$ of the coupling 
constant.  If $g_*$ is small, a perturbative study of
the infrared behavior of the theory might be warranted, but
for larger values of $g_*$ the investigation must proceed
through non-perturbative techniques.

In the specific case
of $SU(3)$, lattice studies indicate that the 12 fundamental flavor theory
exhibits an infrared fixed point~\cite{Appelquist:2011dp,Hasenfratz:2011xn,Aoki:2012eq,DeGrand:2011cu,Cheng:2013xha,Cheng:2014jba,Lombardo:2014pda,Aoki:2013zsa}. Moreover recent
lattice investigations of the $SU(3)$ theory with 8 fundamental,  and 2 sextet flavors have given tantalizing evidence for the existence of a
low mass scalar~\cite{Aoki:2014xpa,Fodor:2014pqa}, as would be needed for a model that describes
the Higgs. 

While the above considerations may offer some hope that the 
explanation of electroweak symmetry breaking could be found in a
strongly interacting gauge theory, an enormous amount of work
remains to be done  to support or invalidate such a conjecture.  Indeed, beyond the intrinsic difficulty
of calculating observables of a non-Abelian gauge theory in
the intermediate coupling domain, a task which can presently be tackled
only by lattice simulation techniques, one also needs to contend
with the fact that, contrary to QCD, the parameters of the
underlying theory (number of colors, number of flavors, 
fermion representation) are not {\it a priori}\/ known, so that one
needs to explore a  range of possible models.  With this paper
we wish to make a contribution to the methodology that can be
used in the search for a successful theory.

\section{\texorpdfstring{Variable $N_f$ vs.~variable mass}{Variable Nf vs. variable mass}}
\label{varmass}

\begin{figure}[tbh]
\begin{center}
    \includegraphics[width=8cm]{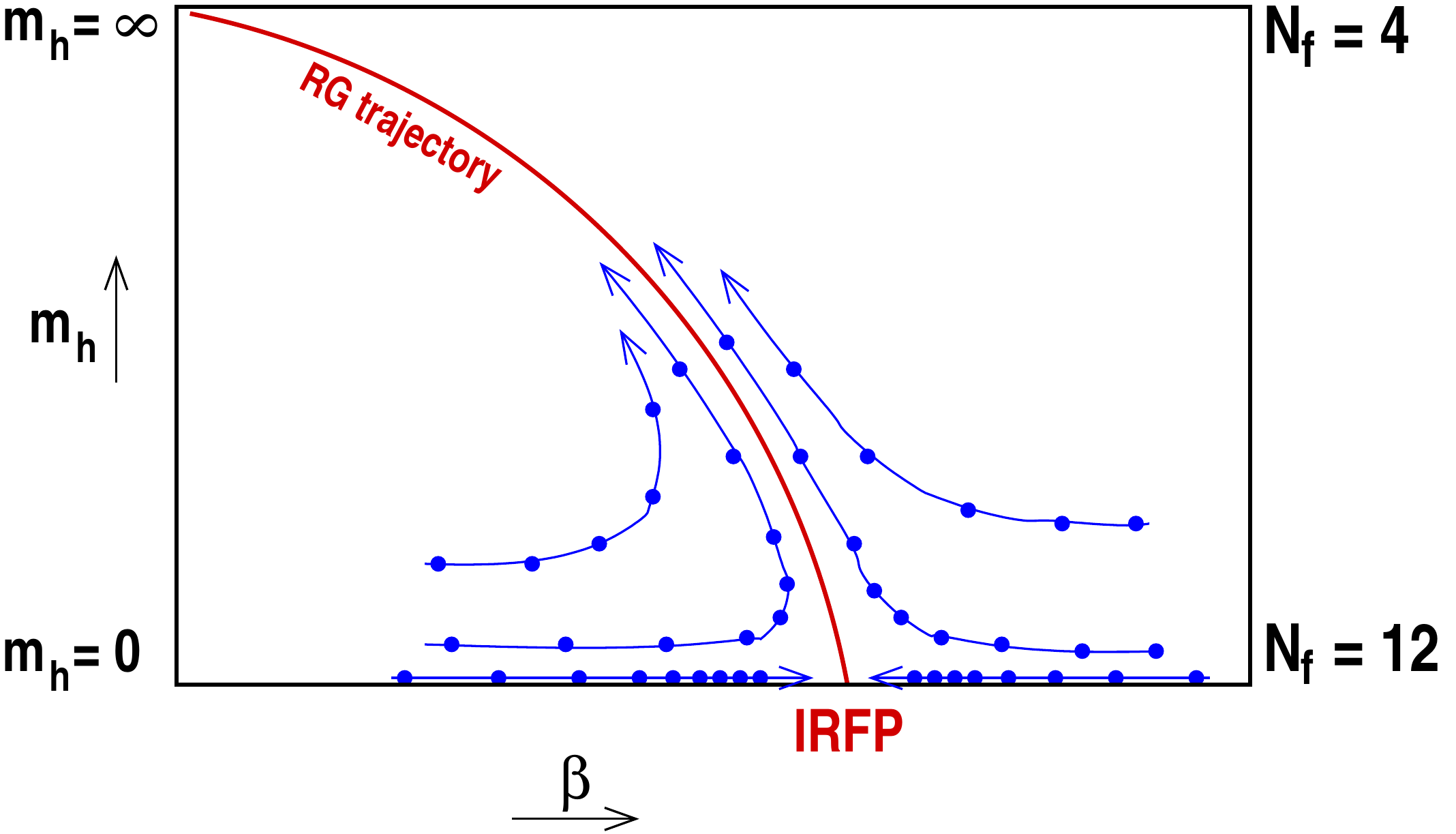}
\end{center}
    \caption{Illustration of the expected renormalization group flow lines  for the $N_\ell+N_h$ flavor theory.  The red line shows the RG trajectory connecting the conformal IRFP at $m_h=m_\ell=0$ (12 flavors) and the trivial fixed point of the 4-flavor theory at $m_h=\infty$. The blue lines are RG flow lines that first approach, then run along  the RG trajectory.  As $m_h \to 0$ the flow lines spend increasingly more time around the IRFP, creating a ``walking'' scenario, while as $m_h$ increases the heavy flavors decouple and the RG flows resemble the running of the  4-flavor system. }  
    \label{f3}
\end{figure}

We consider a system with $N_f$ fermions such that $N_f^{(c)} \le N_f \le N_f'$, i.e.
the theory, while still asymptotic free, has an infrared fixed
point.  One would like to find the value $N_f^{(c)}$ at which
the fixed-point appears. A caveat in this question is that the number of flavors is not a continuous variable, but instead an integer. Further, there is no guarantee that the nearest integer below
$N_f^{(c)}$ is close enough to the infrared fixed point to show the phenomenologically desired properties.

We are suggesting here a different strategy, namely to consider a theory with $N_\ell$ light and $N_h$ heavy flavors. The $N_\ell$ light flavors will be kept near the chiral limit $m_\ell \approx 0$, while the $N_h$ heavy flavors  will have a variable, heavier mass $m_h \ge m_\ell$.
Specifically we study an SU(3) system with  $N_\ell = 4$ and $N_h = 8$ such that the light flavors on their own form a chirally broken system, whereas in the $m_h \rightarrow m_\ell$  limit the $N_\ell+N_h=12$ flavors describe a mass-deformed conformal system with an infrared fixed point. 

To understand the expected behavior as one changes $m_h$, we consider the renormalization group (RG) flow of the $N_\ell+N_h$ theory in the limit $m_\ell=0$, sketched in Figure~\ref{f3}. The heavy fermion mass $m_h$  is a relevant parameter, and for non-zero values it traces out an RG trajectory  leading away from the infra-red fixed point (IRFP) of the $12$-flavor theory to the trivial  fixed point of the $4$-flavor theory. 
For large values of $m_h$, the heavy flavors decouple and the model is essentially the $N_\ell$ flavor chirally broken theory.  In the other limit, when  $m_h=m_\ell=0$, the system is conformal and the RG flow runs into the IRFP. 
Finite  $m_h\ne 0 $ breaks conformal symmetry, but for small $m_h$ the RG flow approaches the IRFP and stays around it for a while before eventually running  toward the trivial infrared fixed point.  This is the desired walking behavior, where the length of walking  can be controlled by tuning $m_h$.
 Thus our theory interpolates between the running behavior of the $4$-flavor chirally broken theory and the walking behavior of the $12$-flavor mass deformed conformal theory.

Phenomenologically it would be more interesting to set $N_\ell=2$ as opposed to 4 to ensure the system has only three massless Goldstone bosons as needed for electroweak symmetry breaking, but due to  our lattice fermion formulation it is simpler in this pilot study to work with four light flavors. As we mentioned earlier, there is increasing evidence that the 12-flavor system is infrared conformal with a relatively small $\gamma_m \approx 0.24$ anomalous dimension~\cite{Cheng:2013eu,Cheng:2013xha,Lombardo:2014pda}, which may be too small to satisfy phenomenological walking constraints. A system closer to the conformal window with a larger anomalous dimension might have been more realistic, but for technical reasons we kept the number of flavors as multiple of four.

\section{Preliminary results}

We have  performed simulations
 at various values of $m_\ell$ with the intention that
the light fermions should be taken in the chiral limit, or
as close as possible to it, while the mass of the heavier fermions 
is varied  with $m_h \ge m_{\ell}$.  
Our calculations are still in progress, the results we present here are preliminary.

\begin{figure}[tb]
{\begin{picture}(135, 55)
   \put(0,1){\includegraphics[width=0.49\textwidth]{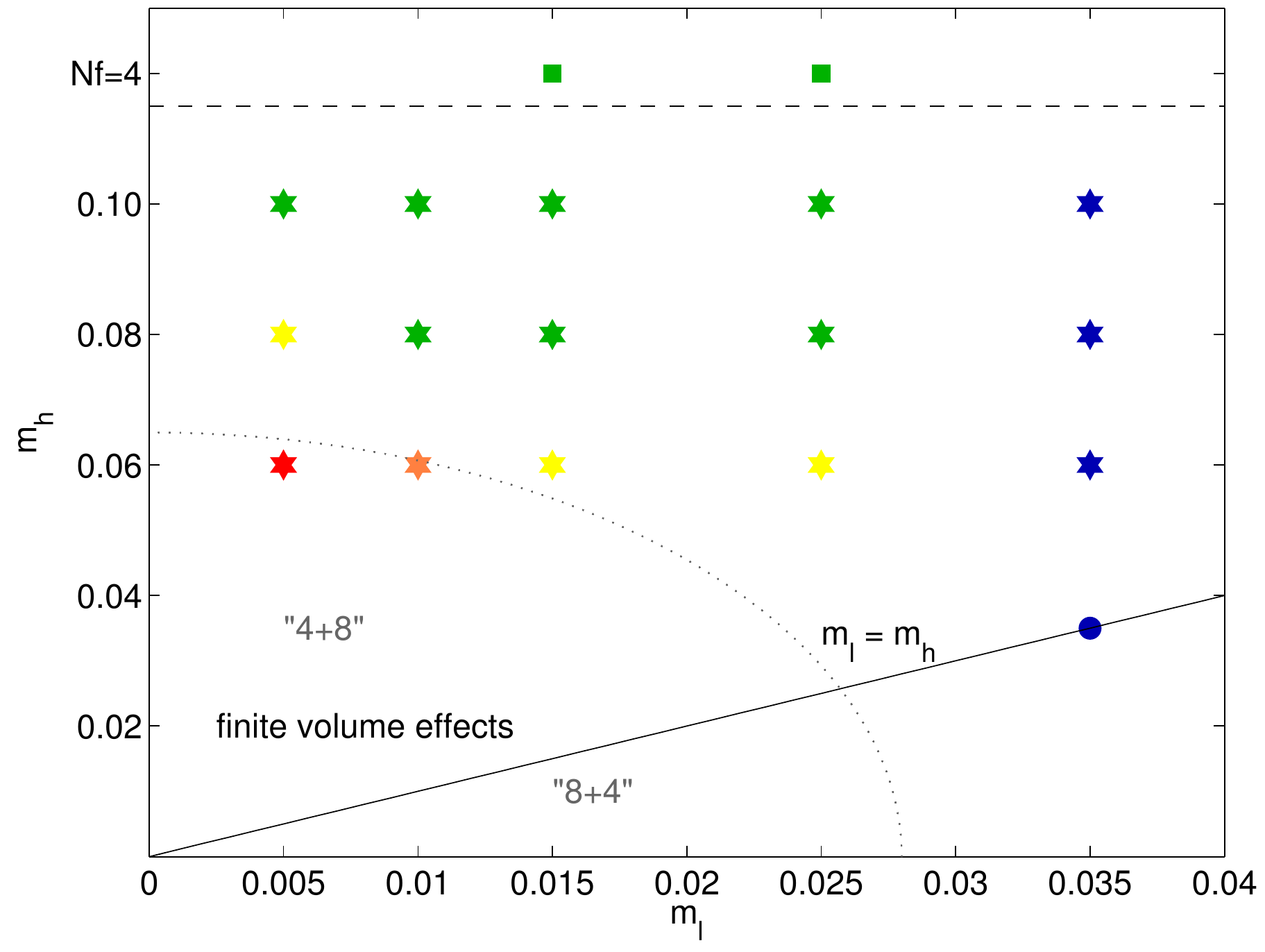}}
   \put(68,0){ \includegraphics[width=0.48\textwidth]{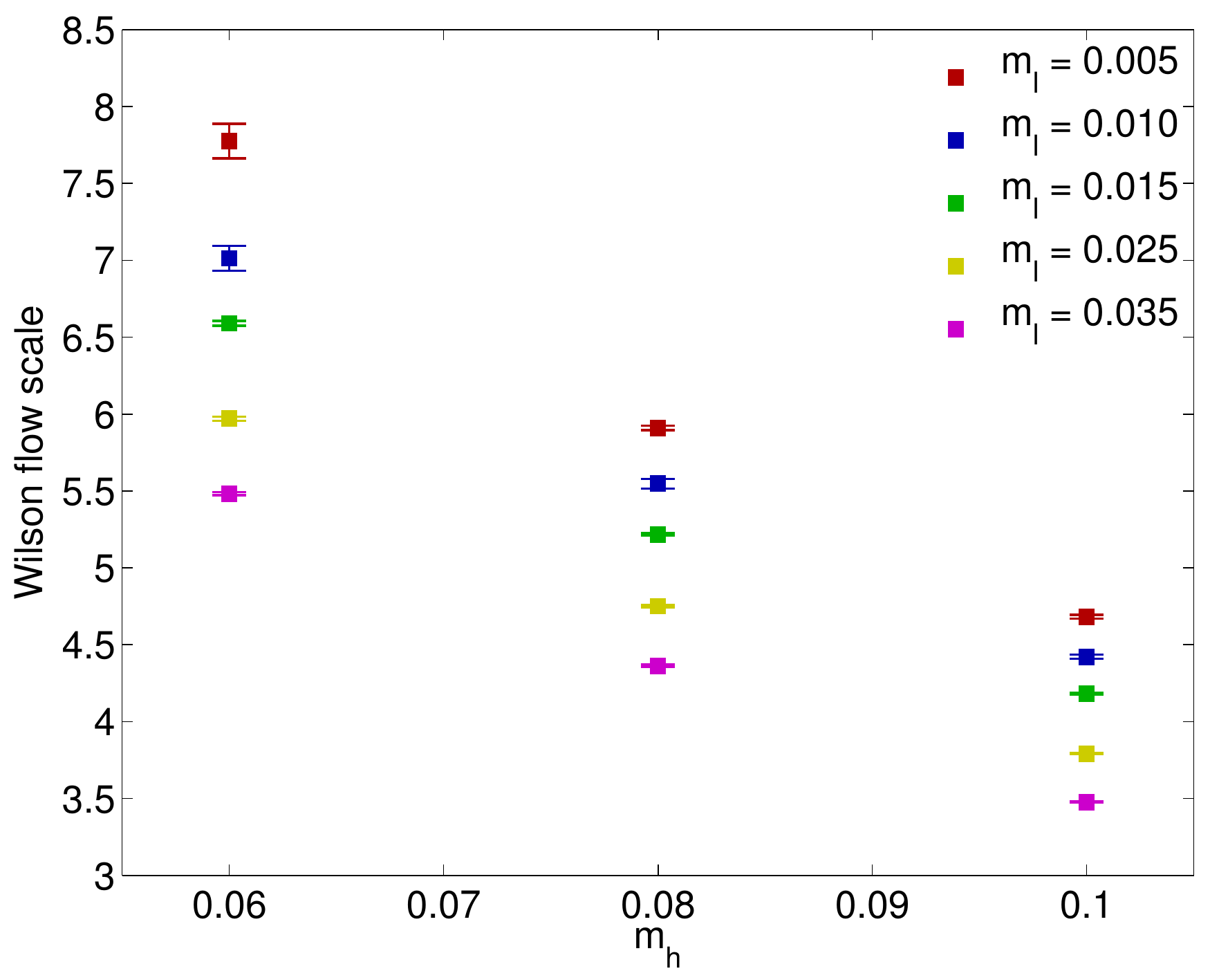}}
   \put(70,37){\rotatebox{90}{\tiny$\mathsf{\sqrt{8t_0}}$}}
\end{picture}}
    \caption{Left panel: light ($m_\ell$) and heavy ($m_h$) mass values for the simulations
carried out on $24^3\times 48$ lattices. The colors are meant to
caution about finite size effects, likely negligible for blue and
green, but of increasing importance as the color turns to yellow, orange
and red. Right panel:  The Wilson flow scale $\sqrt{8t_0}$ for our ensembles. The strong dependence on both the heavy and light fermion masses is most likely due to  the   IRFP of the 12 flavor system.}
    \label{f2}
\end{figure}

The simulations have been carried out using nHYP smeared  staggered fermions
 with fundamental and adjoint plaquette terms  in the gauge action \cite{Cheng:2013xha,Cheng:2013bca} using the FUEL software package \cite{FUEL}.  The technical details will be presented in some further publications.  At present we consider only one gauge coupling, $\beta=4.0$.  
 We performed a large number of simulations on lattices 
of size $24^3\times 48$,
and exploratory runs on lattices of size $32^3 \times 64$.
 The left panel of Figure~\ref{f2} illustrates
the mass values for which we carried out our simulations.

\begin{figure}[tb]
\begin{center}
    \includegraphics[width=8cm]{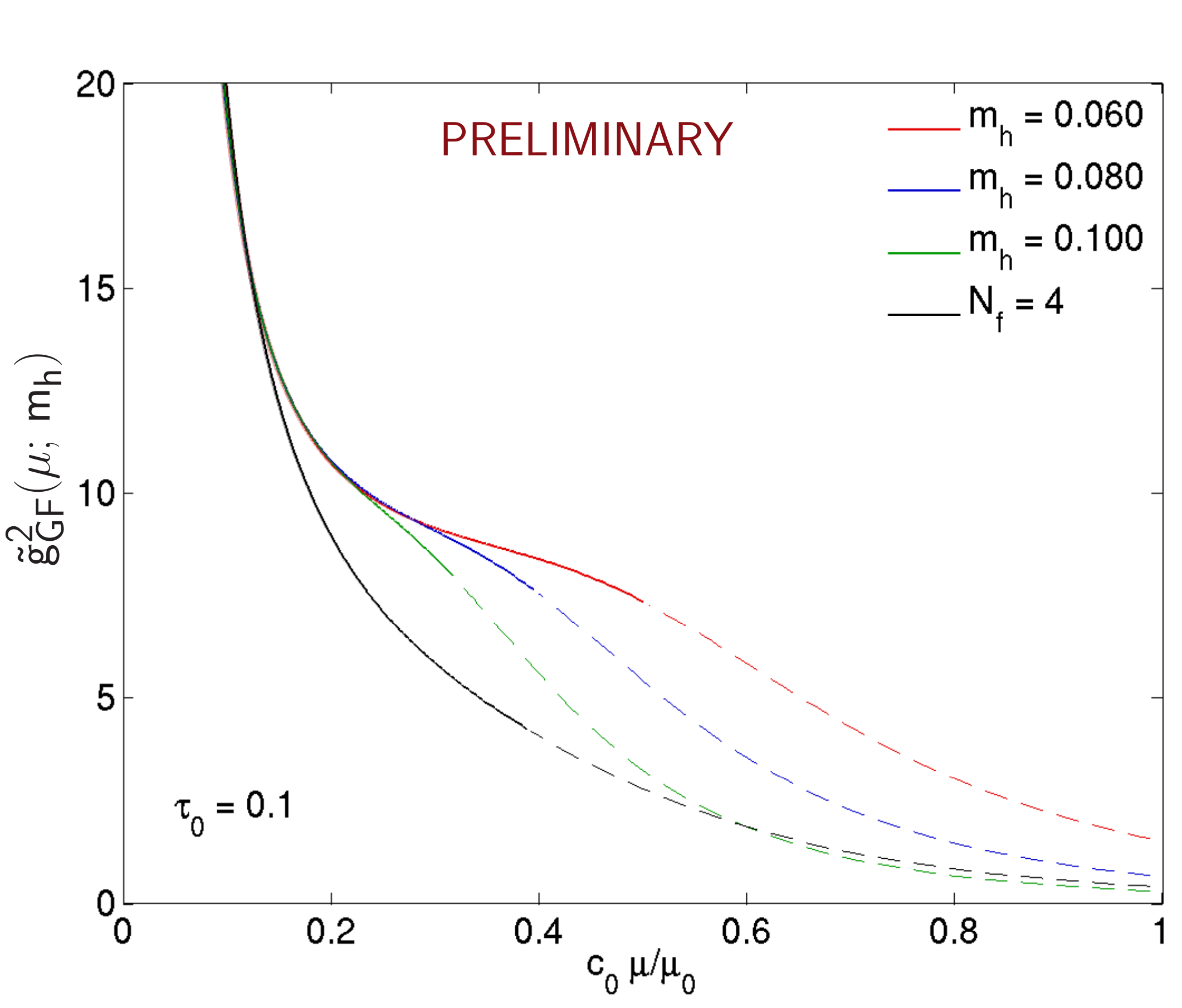} 
\end{center}
    \caption{The running coupling constant $\tilde g_{GF}$ at the mass scale $\mu$ for different values of $m_h$  
      with $m_\ell$ extrapolated to the chiral limit. 
       $\mu_0$ and $c_0$ serve as normalization constants that ensure that the different systems are compared at matching energy scales and $\tau_0$ is the
      shift parameter to remove discretization errors. The dashed portions of the lines indicate where we  suspect cut-off effects  may dominate.}
    \label{f6}
\end{figure}

We use the gradient flow running coupling~\cite{Narayanan:2006rf,Luscher:2009eq,Luscher:2010iy}  to compare the lattice scales of our different ensembles, and also to investigate the energy dependence and walking behavior of the running coupling. The gradient flow is an invertible transformation of the gauge fields that can be used to define a renormalized running coupling at an energy scale $\mu = 1/\sqrt{8t}$ as 
\begin{equation}
g^2_{GF}(\mu) = \frac{1}{\mathcal{N}} \langle t^2 E(t) \rangle ,
\label{eqn8}
\end{equation}
where $t$ is the flow time, 
\begin{equation}
E(t) = -\frac{1}{2}{\rm Re Tr}[G_{\mu\nu}(t)G^{\mu\nu}(t)]
\label{eq7}
\end{equation}
is the energy density, and the constant $\mathcal{N}$ is chosen such that $g^2_{GF}$ matches the traditional $\overline{MS}$ coupling in perturbation theory~\cite{Luscher:2010iy}. The gradient flow scale $t_0$ is defined by fixing the running coupling 
\begin{equation}
g^2_{GF}(t_0) = \frac{0.3}{\mathcal{N}}.
\end{equation}
The right panel of Figure~\ref{f2} shows the quantity $\sqrt{8t_0}$  in lattice units from our $24^3\times48$ volume simulations. To control finite volume effects $\sqrt{8t_0} \lesssim L/5$ is usually sufficient,  but because our model can have stronger finite volume effects  we will  test this conjecture on $32^3\times 64$ volumes.  We also find relatively strong  dependence of $\sqrt{8t_0}$ on the light mass $m_\ell$  indicating that we might need even larger volumes when taking the $m_\ell\to 0$ chiral limit. 

The  gradient flow coupling of Equation~\ref{eqn8} allows us to investigate the energy dependence of the running coupling.
By implementing a simple  modification  to the gradient flow coupling 
\begin{equation}
\tilde g^2_{GF}(\mu) = \frac{1}{\mathcal{N}} \langle t^2 E(t+a^2 \tau_0) \rangle ,
\label{eqn9}
\end{equation}
where $a^2 \tau_0 \ll t$ is a small shift in the flow time, we can remove most of the   discretization errors and  improve the convergence behavior of the renormalized coupling toward the continuum limit~\cite{Cheng:2014jba,AnnaLat14Talk}.  Figure~\ref{f6} shows the running coupling as function of the scale parameter for different values of $m_h$ and for the four flavor theory, equivalent to the limit $m_h \to \infty$.

Our results shown in Fig.~\ref{f6}, albeit preliminary,
indicate that as the heavy fermion mass is reduced and
 the conformal limit is approached, the running of the coupling
constant slows down, as one would need for an application
of strong dynamics models to electroweak symmetry breaking.

\begin{figure}[!ht]
\begin{center}
    \includegraphics[width=8cm]{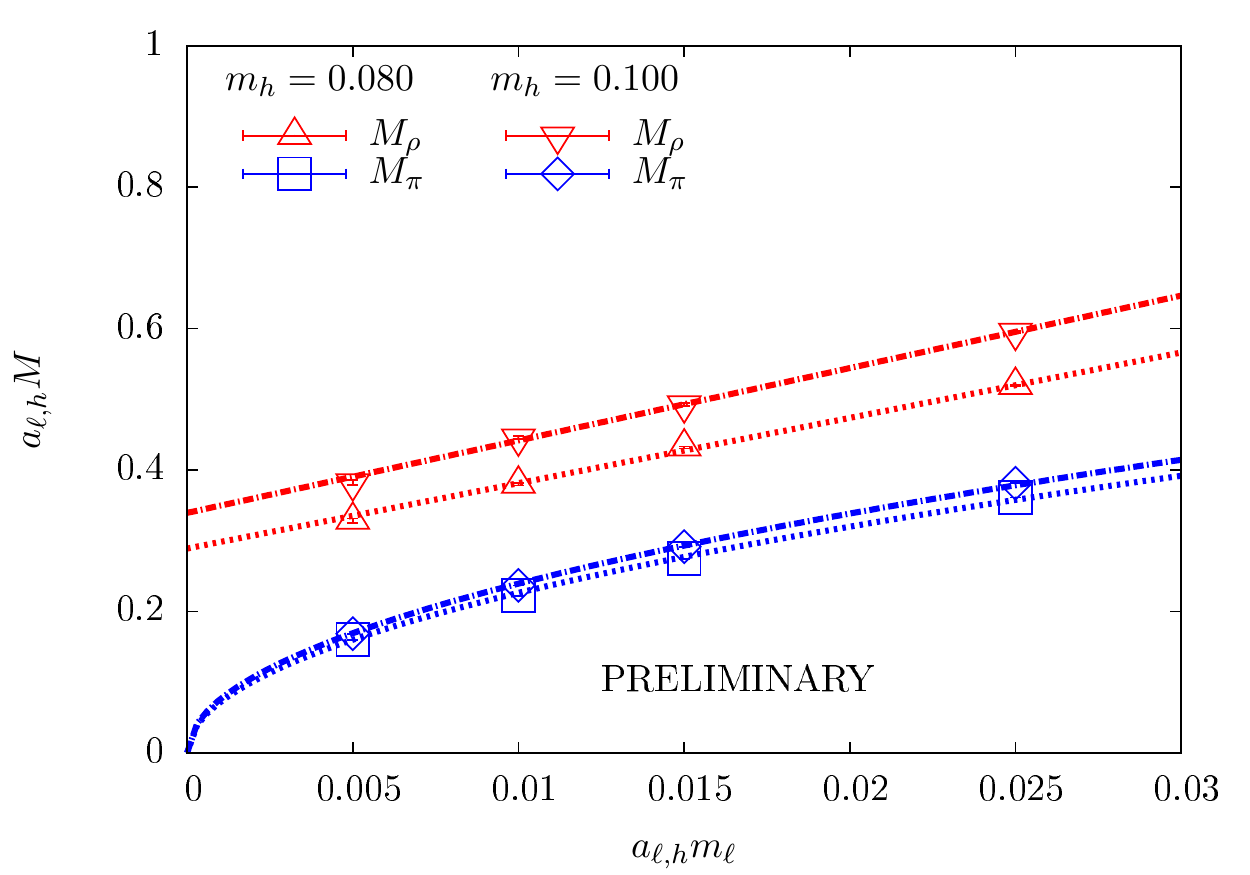} 
\end{center}
    \caption{Masses of the Goldstone boson pion and the rho meson in lattice units for ensembles with $m_h = 0.080$ and $0.100$. The dashed lines serve to guide the eye indicating the expected extrapolation: linearly for the masses of the rho-meson and proportional to $\sqrt{m_\ell}$ for the pion masses.}
    \label{fig7}
\end{figure}

Finally, we present first results for the connected meson spectrum of the light flavors. We extract the meson masses from a correlated fit to 2-point correlators obtained from measurements performed with wall sources in order to suppress couplings to excited or scattering states~\cite{Gupta:1990mr}.  In Figure \ref{fig7} we show our  results for the Goldstone boson pion and the rho meson obtained for  two values of the heavy mass, $m_h = 0.080$ and $m_h = 0.100$. The  data are shown in bare lattice units without taking into account  the relatively large changes of the lattice scale when varying $m_\ell$ and $m_h$.  

The very interesting $0^{++}$ scalar meson mass is much harder to extract as it requires the evaluation of disconnected diagrams and  needs significantly higher statistics. Our preliminary results indicate that our present statistics is sufficient to predict the mass of the scalar and we expect to finalize and present our results in the near future.

\section{Conclusions}
\label{conclusions}
Gauge-fermion systems near the conformal window are interesting not only as strongly coupled field theory models but could also have
important phenomenological applications. In recent years many large-scale lattice studies  have began to investigate systems with different gauge groups, fermion representations, and fermion numbers with the goal of identifying the onset of conformal behavior and investigating the properties of systems just below the conformal window.
In the present  work, we propose to study the transition from conformal behavior with an infrared
fixed-point to chiral symmetry breaking.  Starting with a conformal system, we lift the mass of some of the fermions  so that they decouple in the infrared limit. If the remaining fermions  are near the chiral limit, this allows a controlled transition between conformality and chiral symmetry breaking. A walking gauge coupling emerges naturally in this system.  
 In this article we presented preliminary results supporting our expectations. If further investigations
confirm  our preliminary results, our model may play an important role for the
study of theories of electroweak symmetry breaking based
on strong dynamics. 

\section{Acknowledgements}
\label{acknowledgements}

Computations for this work were carried out in part on facilities of
the USQCD Collaboration, which are funded by the Office of Science of the U.S. Department of Energy, on computers at the MGHPCC, in part funded by the National Science Foundation, and on computers allocated under the NSF Xsede program to the project TG-PHY120002. \\
We thank Boston University, Fermilab, the NSF and the U.S. DOE for providing the facilities essential for the completion of this work.
R.C.B., C.R. and E.W. were supported by DOE grant DE-SC0010025. In addition, R.C.B., C.R. and O.W. acknowledge the support of NSF grant OCI-0749300. A.H. acknowledges support by the DOE grant DE-SC0010005.

\bibliography{../General/BSM}
\bibliographystyle{apsrev4-1}

\end{document}